\documentclass[transactions,a4paper]{IEEEtran}

\usepackage{color}
\definecolor{gray}{rgb}{0.5,0.5,0.5}
\newcommand{\added}[1]{#1}
\newcommand{\changed}[1]{#1}

\newcommand{\addedold}[1]{#1}
\newcommand{\changedold}[1]{#1}
\newcommand{\changedtwo}[1]{#1}

\usepackage[
pdftitle={Stability of Evolving Multi-Agent Systems},
pdfauthor={Gerard Briscoe},
pdfsubject={Multi-Agent Systems},
pdfkeywords={evolution, stability, agent, entropy, equilibrium},
colorlinks=true, linkcolor=black, citecolor=black, filecolor=black, urlcolor=black, hyperfootnotes=false]{hyperref}

\usepackage{graphicx, acronym, ifthen, substr, color}
\usepackage{amsthm, amsmath}

\newboolean{final}\setboolean{final}{true}
\newboolean{mactex}\setboolean{mactex}{true}
\newboolean{arxiv}\setboolean{arxiv}{false}

\ifthenelse{\boolean{final}}{}{\usepackage{showlabels}\usepackage{citeext}}

\ifthenelse{\boolean{arxiv}}
    {\newcommand{\arxivfootnote}{\footnote}}
    {\def\arxivfootnote#1{}}

\ifthenelse{\boolean{arxiv}}
    {\def\arxivOnly#1{#1}}
    {\def\arxivOnly#1{}}

\ifthenelse{\boolean{arxiv}}
    {\def\arxivNot#1{}}
    {\def\arxivNot#1{#1}}

\ifthenelse{\boolean{arxiv}}
    {}
    {}

\definecolor{tred}{RGB}{255,0,0}

\newcommand{\tfigure}[9]
    {
    \IfSubStringInString{!}{#7}{\begin{figure}[#7]}{\begin{figure}[!t]}
    \IfSubStringInString{mm}{#8}{\vspace{#8}}{}
    \centering

    \IfSubStringInString{pdf}{#3}
        {
        \includegraphics[#1]{images/#2}
        }
        {
        \execute{cd images; ./pdfcrop.sh #2}
        \includegraphics[#1]{images/#2-crop.pdf}
        }
    \vspace{#6}
    \caption[#4]
        {
        \label{#2}
        #4: #5
        }
    \IfSubStringInString{mm}{#9}{\vspace{#9}}{}
    \end{figure}
    }

\newcommand{\Circlesub}[4]
    {
    \ifthenelse{\boolean{mactex}}{}{\immediate\write18{cd images; ./pdfcrop.sh circle#2}}
    \ifthenelse{\boolean{final}}
        {\hspace{#1}\raisebox{#4}{$\includegraphics[scale=0.5, clip=true, trim=0mm 0mm 0mm 0mm]{images/circle#2-crop.pdf}$}\hspace{#3}}
        {\href{file://localhost/Users/g/Desktop/PhDthesis/images/circle#2.graffle}{\hspace{#1}\raisebox{#4}{$\includegraphics[clip=true, trim=0mm 0mm 0mm 0mm]{images/circle#2-crop.pdf}$}\hspace{#3}}}
    }

\newcommand{\execute}[1]{\immediate\write18{#1}}

\newcommand{\setCap}[2]{#1\immediate\write18{./mkcaption.sh #2}}
\newcommand{\getCap}[1]{\acl*{#1}}

\acrodef{PCG}{Projected Conjugate Gradient} 
\acrodef{QP}{quadratic programming}
\acrodef{RBF}{Radial-Basis Function}
\acrodef{ABM}{Agent-Based Modelling}
\acrodef{AI}{Artificial Intelligence}
\acrodef{DAI}{Distributed Artificial Intelligence}
\acrodef{API}{Application Programming Interface}
\acrodef{ARF}{p14ARF human tumor-suppressor gene}
\acrodef{B2B}{business-to-business}
\acrodef{BDP}{Biological Design Pattern}
\acrodef{BGS}{Best Guess Solution}
\acrodef{BIC}{Biologically-Inspired Computing}
\acrodef{BML}{Business Modelling Language}
\acrodef{BPEL}{Business Process Execution Language}
\acrodef{BPMN}{Business Process Modelling Notation}
\acrodef{CAS}{Complex Adaptive Systems}
\acrodef{COBOL}{COmmon Business-Oriented Language}
\acrodef{DBE}{Digital Business Ecosystem}
\acrodef{DE}{Digital Ecosystem}
\acrodef{DEC}{distributed evolutionary computing}
\acrodef{DGA}{Distributed genetic algorithms}
\acrodef{DIS}{Distributed Intelligence System}
\acrodef{DNA}{Deoxyribose Nucleic Acid}
\acrodef{DOP}{DBE Open Protocol}
\acrodef{DSS}{Distributed Storage System}
\acrodef{EAP}{Evolving Agent Population}
\acrodef{ebXML}{e-business eXtensible Markup Language}
\acrodef{EC}{Evolutionary Computing}
\acrodef{ECJ}{Evolutionary Computing in Java}
\acrodef{EE}{Evolutionary Environment}
\acrodef{EFL}{Evolutionary Framework for Language}
\acrodef{FLE}{Framework for Language Ecosystems}
\acrodef{EOA}{Ecosystem-Oriented Architecture}
\acrodef{ESS}{evolutionary stable strategy}
\acrodef{EvE}{Evolutionary Environment}
\acrodef{ExE}{Execution Environment}
\acrodef{FCB}{Framework for Computational Biomimicry}
\acrodef{FFF}{Fitness Function Framework}
\acrodef{FL}{Fitness Landscape}
\acrodef{HWU}{Heriot-Watt University}
\acrodef{ICL}{Imperial College London}
\acrodef{ICT}{Information and Communications Technology}
\acrodef{INTEL}{Intel Ireland}
\acrodef{IPA}{International Phonetic Alphabet}
\acrodef{ISUFI}{Istituto Superiore Universitario di Formazione Interdisciplinare}
\acrodef{JDJ}{Java Developer's Journal}
\acrodef{KC}{Kolmogorov-Chaitin}
\acrodef{LAN}{local area network}
\acrodef{LSE}{London School of Economics and Political Science}
\acrodef{MAS}{Multi-Agent System}
\acrodef{MDL}{Minimum Description Length}
\acrodef{MDM2}{murine double minute 2}
\acrodef{MFT}{Mean Field Theory}
\acrodef{MoAS}{Mobile Agent System}
\acrodef{MOF}{Meta Object Facility}
\acrodef{MUH}{migration and usage history}
\acrodef{NIC}{Nature Inspired Computing}
\acrodef{NN}{Neural Network}
\acrodef{NoE}{Network of Excellence}
\acrodef{OMG}{Open Mac Grid}
\acrodef{OPAALS}{Open Philosophies for Associative Autopoietic Digital Ecosystems}
\acrodef{P2P}{peer-to-peer}
\acrodef{P53}{protein 53}
\acrodef{PDA}{Personal Digital Assistant}
\acrodef{QoS}{quality of service}
\acrodef{REST}{REpresentational State Transfer}
\acrodef{RNA}{Deoxyribose Nucleic Acid}
\acrodef{SAE}{Software Agent Ecosystem}
\acrodef{SBML}{Systems Biology Modelling Language}
\acrodef{SBVR}{Semantics of Business Vocabulary and Business Rules}
\acrodef{SDL}{Service Description Language}
\acrodef{SF}{Service Factory}
\acrodef{SIM}{Social Interaction Mechanism}
\acrodef{SM}{Service Manifest}
\acrodef{SME}{Small and Medium sized Enterprise}
\acrodef{SML}{Service Modelling Language}
\acrodef{SMO}{Sequential Minimal Optimisation}
\acrodef{SOA}{Service-Oriented Architecture}
\acrodef{SOAP}{Simple Object Access Protocol}
\acrodef{SOC}{Self-Organised Criticality}
\acrodef{SOLUTA}{SOLUTA.NET}
\acrodef{SOM}{Self-Organising Map}
\acrodef{SSL}{Semantic Service Language}
\acrodef{STU}{Salzburg Technical University}
\acrodef{SUN}{Sun Microsystems}
\acrodef{SVM}{Support Vector Machine}
\acrodef{TM}{Turing Machine}
\acrodef{UBHAM}{University of Birmingham}
\acrodef{UDDI}{Universal Description Discovery and Integration}
\acrodef{UML}{Unified Modelling Language}
\acrodef{URI}{Uniform Resource Identifier}
\acrodef{UTM}{Universal Turing Machine}
\acrodef{VLP}{variable length population}
\acrodef{VLS}{variable length sequences}
\acrodef{vls}{variable length sequence}
\acrodef{WP}{Work-Package}
\acrodef{WSDL}{Web Services Definition Language}
\acrodef{XMI}{XML Metadata Interchange}
\acrodef{XML}{eXtensible Markup Language}
\acrodef{MD5}{Message-Digest algorithm 5}
\acrodef{GA}{genetic algorithm}
\acrodef{GP}{genetic programming}
\acrodef{MASON}{Multi-Agent Simulator Of Neighbourhoods}
\acrodef{Repast}{Recursive Porous Agent Simulation Toolkit}
\acrodef{JCLEC}{Java Computing Library for Evolutionary Computing}
\acrodef{OWL-S}{Web Ontology Language - Service}
\acrodef{EGT}{Evolutionary Game Theory}
\acrodef{RBF}{Radial Basis Functions}
\acrodef{SWS}{Semantic Web Services}
\acrodef{HDD}{Hard Disk Drive}
\acrodef{SSD}{Solid-State Drive}

\acrodef{statesCap}{possible evolutionary path through the state-space}
\acrodef{capStates3}{the {selection pressure} of the evolutionary process}
\acrodef{capStates}{driving it towards the {maximal state} of the maximum macro-state $M_{max}$}
\acrodef{as3}{with an abstract representation consisting of a set of}
\acrodef{agentSemantic2}{attribute representing a {property} of the {semantic description}, ranging between one and a hundred.}
\acrodef{as4}{Each simulated agent was initialised with a semantic description}
\acrodef{evoGraph}{shows both the maximum and average fitness increasing over the generations of a typical evolving agent population, and as expected the {average fitness} remains below the {maximum fitness} because of variation in the evolving agent population \cite{goldberg}}
\acrodef{graphCap}{in the maximum macro-state $M_{max}$ only after generation 178 and always after generation 482. It was also observed being in the sub-optimal macro-state $M_{half}$ only between generations 37 and 113}
\acrodef{aScap}{With the mutation rate under or equal to 60\%, the evolving agent population showed no instability, with $\delta$ values equal to zero as the system $S$ was always in the same macro-state $M$ at infinite time, independent of the crossover rate. With the mutation rate above 60\% the instability increased significantly}

\immediate\write18{echo > captions.tex}

\newtheorem{definition}{Definition}
\newtheorem{theorem}{Theorem}

\newcommand{\be}{\begin{equation}}
\newcommand{\eeq}[1]{\label{#1}\end{equation}}
\newcommand{\bx}{{\bf X}}
\newcommand{\by}{{\bf Y}}
\newcommand{\bone}{{\bf M_{max}}}
\newcommand{\bfif}{{\bf M_{half}}}

\newcommand{\bxi}{\mbox{\boldmath $\xi$}}
\newcommand{\ra}{1, \ldots, n}

\begin{document}

\title{Stability of Evolving Multi-Agent Systems}

\author{Philippe~De~Wilde,~\IEEEmembership{Senior Member,~IEEE}, and Gerard~Briscoe
\thanks{Philippe De Wilde is with the Intelligent Systems Lab, Department of Computer Science, Heriot Watt University, United Kingdom, pdw@macs.hw.ac.uk.}
\thanks{Gerard Briscoe is with the Systems Research Group, Computer Laboratory, University of Cambridge, United Kingdom, gerard.briscoe@cl.cam.ac.uk.}
}

\maketitle

\begin{abstract}
A \acl{MAS} is a distributed system where the agents or nodes perform complex functions that cannot be written down in analytic form. \aclp{MAS} are highly connected, and the information they contain is mostly stored in the connections. When agents update their state, they take into account the state of the other agents, and they have access to those states via the connections. There is also external, user-generated input into the \acl{MAS}. As so much information is stored in the connections, agents are often memory-less. This memory-less property, together with the randomness of the external input, has allowed us to model \aclp{MAS} using Markov chains. In this paper, we look at \aclp{MAS} that evolve, i.e. the number of agents varies according to the fitness of the individual agents. We extend our Markov chain model, and define {\em stability}. This is the start of a methodology to {\em control} \aclp{MAS}. We then build upon this to construct an entropy-based definition for the \emph{degree of instability} (entropy of the limit probabilities), which we used to perform a \emph{stability analysis}. We then investigated the stability of evolving agent populations through simulation, and show that the results are consistent with the original definition of stability in non-evolving \aclp{MAS}, proposed by Chli and De Wilde. This paper forms the theoretical basis for the construction of Digital Business Ecosystems, and applications have been reported elsewhere.

\end{abstract}

\IEEEpeerreviewmaketitle

\section{Introduction}

\aclp{MAS} is a growing field primarily because of recent developments of the Internet as a means of circulating information, goods and services. Many researchers have contributed valuable work to the area in recent years \cite{nwana1996software}. However, despite both Evolutionary Computing and \aclp{MAS} being mature research areas \cite{masOverviewPaper, ecpaper} their integration, creating evolving agent populations, is a recent development \cite{smith1998fec}. This integration is non-trivial, because agents can be considered as state machines and Evolutionary Computing algorithms have been developed to work on numerical data and strings without memory effects. So, our aim here is to determine, for \aclp{MAS} which make use of Evolutionary Computing \cite{mabu2007gbe, bionetics, smith2000eec}, macroscopic variables that characterise their stability.

\changedold{While there are several definitions of stability defined for \aclp{MAS} \addedold{\cite{olfatisaber2007cac, angeli2006stability, moreau2005sms, weiss1999msm, mohanarajah2008formation}}, they are not applicable because of the Evolutionary Computing dynamics inherent in the context of evolving agent populations. Chli and De Wilde \cite{chli2} model \aclp{MAS} as Markov chains, which are an established modelling approach in Evolutionary Computing \cite{rudolph1998fmc}. They model agent evolution in time as Markov processes, and so view a \acl{MAS} as a discrete time Markov chain with potentially unknown transition probabilities, considered \emph{stable} when its state has converged to an equilibrium distribution \cite{chli2}. Also, while there is past work on modelling Evolutionary Computing algorithms as Markov chains \cite{rudolph, nix, goldberg2, eibenAarts}, we have found none including \aclp{MAS}.}

\changedold{Therefore, we decided to \emph{extend} the \addedold{\emph{existing}} Chli-DeWilde definition of agent stability to include the dynamics of Evolutionary Computing, \addedold{including \emph{population dynamics} and \emph{macro-states} of the population state-space.}}

\changed{We have applied our efforts reported here to the construction of Digital Business Ecosystems \cite{dbebook, caes}, Digital Ecosystems (distributed adaptive open socio-technical systems, with properties of self-organisation, scalability and sustainability, inspired by natural ecosystems) for conducting business that enable network-based economies, levelling the playing field for \acp{SME} \cite{thesis, epi}. \acp{SME} provide substantial employment and conduct much innovative activity, but struggle in global markets on a far from level playing field, where large companies have distinct advantages \cite{abcdbe}. So, we created Digital Ecosystems to be the \emph{digital counterparts of natural ecosystems} \cite{de07oz, acmMedes}, which are considered to be robust, self-organising and scalable architectures that can automatically solve complex, dynamic problems \cite{dbebkpub, thesis, bionetics}. This lead to a novel optimisation technique inspired by natural ecosystems, where the optimisation works at two levels: a first optimisation, migration of agents which are distributed in a decentralised peer-to-peer network, operating continuously in time; this process feeds a second optimisation based on evolutionary computing that operates locally on single peers and is aimed at finding solutions to satisfy locally relevant constraints.} \added{So, the local search is improved through this twofold process to yield better local optima faster, as the distributed optimisation provides prior sampling of the search space through computations already performed in other peers with similar constraints.} \changed{Therefore, our extended Chli-DeWilde stability was invaluable in understanding and developing the evolving agent populations in these \acp{EOA} \cite{agentStability},} \added{because it was important for us to be able to understand, model, and define stability, including the determination of macroscopic variables to characterise the stability, of the order constructing processes within, the evolving agent populations.}

\addedold{\section{Multi-Agent Systems}}

\changedold{A \emph{software agent} is a piece of software that acts, for a user in a relationship of \emph{agency}, autonomously in an environment to meet its designed objectives \cite{wooldridge}. So, a \acl{MAS} is a system composed of several \emph{software agents}, collectively capable of reaching goals that are difficult to achieve by an individual agent or monolithic system \cite{wooldridge}.} \addedold{Examples of problems which are appropriate to \aclp{MAS} research include online trading \cite{rogers2007effects}, disaster response \cite{schurr2005future}, and modelling social structures \cite{sun2004simulating}. Multi-agent systems are applied in the real world to graphical applications such as computer games and films. They are also used for coordinated defence systems, with other applications including transportation, logistics, as well as in many other fields. It is widely being advocated for use in networking and mobile technologies, to achieve automatic and dynamic load balancing, high scalability, and self-healing networks.}

\changedold{The systems studied by Wiener \cite{wiener1948cybernetics} model information flow between an actor and the environment. Input, state and output and defined, and consequently an evolution equation can be established. In such a system it is possible to distinguish an observational sequence (of the inputs), followed by a decisional sequence of outputs \cite{vallee1995cognition}.}

\vspace{2\baselineskip}

\section{Chli-DeWilde Stability}

\addedold{We will now briefly introduce Chli-DeWilde stability for \acl{MAS} and Evolutionary Computing, sufficiently to allow for the derivation of our extensions to Chli-DeWilde stability to include \aclp{MAS} with Evolutionary Computing.} Chli-DeWilde stability was created to provide a clear notion of stability in \aclp{MAS} \cite{chli2}, because while computer scientists often talk about stable or unstable systems \cite{mspaper5ThomasSycara1998, mspaper9Balakrishnan1997}, they did so without having a concrete or uniform definition of stability. So, the Chli-DeWilde definition of stability for \aclp{MAS} was created \cite{chli2}, derived \cite{chlithesis} from the notion of stability defined by De Wilde \cite{mspaperDeWilde1999a, mspaperLee1998}, based on the stationary distribution of a stochastic system, making use of discrete-time Markov chains, which we will now introduce\footnote{A more comprehensive introduction to Markov chain theory and stochastic processes is available in \cite{msthesisNorris1997} and \cite{msthesisCoxMiller1972}.}.

If we let $I$ be a \emph{countable set}, in which each $i \in I$ is called a \emph{state} and $I$ is called the \emph{state-space}. We can then say that $\lambda = (\lambda_i : i \in I)$ is a \emph{measure on} $I$ if $0 \le \lambda_i < \infty$ for all $i \in I$, and additionally a \emph{distribution} if $\sum_{i \in I}{\lambda_i=1}$ \cite{chlithesis}. So, if $X$ is a \emph{random variable} taking values in $I$ and we have $\lambda_i = \Pr(X = i)$, then $\lambda$ is \emph{the distribution of $X$}, and we can say that a matrix $P = (p_{ij} : i,j \in I)$ is \emph{stochastic} if every row $(p_{ij} : j \in I)$ is a \emph{distribution} \changedold{\cite{chlithesis,msthesisNorris1997}}. 
We can then extend familiar notions of matrix and vector multiplication to cover a general index set $I$ of potentially infinite size, by defining the multiplication of a matrix by a measure as $\lambda P$, which is given by
\begin{equation}
(\lambda P)_i = \sum\limits_{j \in I}{\lambda_{j}p_{ij}}.
\label{ms3dot1}
\end{equation}
We can now describe the rules for a Markov chain by a definition in terms of the corresponding matrix $P$ \addedold{\cite{chlithesis,msthesisNorris1997}}.
\\

\begin{definition}
We say that $(X^t)_{t\ge0}$ is a Markov chain with initial distribution $\lambda = (\lambda_i : i \in I)$ and transition matrix $P = (p_{ij} : i,j \in I)$ if:
\begin{enumerate}
\item $\Pr(X^0 = i_0) = \lambda_{i_0}$ and
\item $\Pr(X^{t+1} = i_{t+1}\ |\ X^0 = i_0, \ldots, X^t = i_t) = p_{i_t i_{t+1}}$.
\end{enumerate}
We abbreviate these two conditions by saying that $(X^t)_{t\ge0}$ is Markov$(\lambda, P)$.\\
\end{definition}

From this first definition the Markov process is \emph{memoryless}\addedold{\footnote{\addedold{Markov systems with probabilities is a very powerful modelling technique, applicable in large variety of scenarios, and it is common to start memoryless, in which the output probability distribution only depends on the current input. However, there are scenarios in which alternative modelling techniques, like queueing systems, are more suitable, such as when there is asynchronous communications, and to fully characterise the system state at time (t), the history of states at (t-1), (t-2), ... might also need to be considered.}}}, resulting in only the current state of the system being required to describe its subsequent behaviour. We say that a Markov process $X^0, X^1, \ldots, X^t$ has a \emph{stationary distribution} if the probability distribution of $X^t$ becomes independent of the time $t$ \cite{chli2}. So, the following theorem is an easy consequence of the second condition from the first definition.\\

\begin{theorem}
A discrete-time random process $(X^t)_{t\ge0}$ is Markov$(\lambda,P)$, if and only if for all $t$ and $i_0, \ldots, i_t$ we have
\begin{equation}
\Pr(X^0 = i_0, \ldots, X^t = i_t) = \lambda_{i_0}p_{i_0 i_1} \cdots p_{i_{t-1}i_t}.
\label{ms3dot2}
\end{equation}
\end{theorem}

\vspace{5mm}

This first theorem depicts the structure of a Markov chain \addedold{\cite{chlithesis,msthesisNorris1997,msthesisCoxMiller1972}, illustrating the relation with the stochastic matrix $P$. The next Theorem shows how the Markov chain evolves in time, again showing the role of the matrix $P$.}
\\

\begin{theorem}
Let $(X^t)_{t\ge0}$ be $Markov(\lambda,P)$, then for all $t,s\ge0$:
\begin{enumerate}
\item $\Pr(X^t = j) = (\lambda P^t)_j$ and
\item $\Pr(X^t = j\ |\ X^0 = i) = \Pr(X^{t+s} = j\ |\ X^s = i) = (P^t)_{ij}$.
\end{enumerate}
\label{ms3dot3dot2}
For convenience $(P^t)_{ij}$ can be denoted as $p^{(t)}_{ij}$.
\end{theorem}

\vspace{5mm}

Given this second theorem we can define $p^{(t)}_{ij}$ as the $t$-step transition probability from the state $i$ to $j$ \cite{chlithesis}, and we can now introduce the concept of an \emph{invariant distribution} \cite{chlithesis}, in which we say that $\lambda$ is invariant if
\begin{equation}
\lambda P = \lambda .
\end{equation}
The next theorem will link the existence of an \emph{invariant distribution}, which is an algebraic property of the matrix $P$, with the probabilistic concept of an \emph{equilibrium distribution}. This only applies to a restricted class of Markov chains, namely, those with \emph{irreducible} and \emph{aperiodic} stochastic matrices. However, there is a multitude of analogous results for other types of Markov chains which we can refer \cite{msthesisNorris1997, msthesisCoxMiller1972}, and the following theorem is provided as an indication, of the family of theorems that apply. An \emph{irreducible} matrix $P$ is one for which, for all $i,j \in I$ there exist a sufficiently large $t$ such that $p^{(t)}_{ij} > 0$. I matrix $P$ and is \emph{aperiodic} if for all states $i \in I$ we have $p^{(t)}_{ii} > 0$ for all sufficiently large $t$ \addedold{\cite{chlithesis,msthesisNorris1997,msthesisCoxMiller1972}}. \addedold{The meaning of these properties can broadly be explained as follows. An irreducible Markov chain is a chain where all states intercommunicate. For this to happen, there needs to be a non-zero probability to go from any state to any other state. This communication can happen in any number $t$ of time steps. This leads to the condition $p^{(t)}_{ij} > 0$ for all $i$ and $j$. An aperiodic Markov chain is a chain where all states are aperiodic. A state is aperiodic if it is not periodic. Finally, a state is periodic if subsequent occupations of this state occur at regular multiples of a time interval. For this to happen, $p^{(t)}_{ii}$ has to be zero for $t$ an integral multiple of a number. This leads to the condition $p^{(t)}_{ii} > 0$ for {\em a}-periodicity. For further explanations, please refer to \cite{msthesisCoxMiller1972}.}\\

\begin{theorem}
Let $P$ be irreducible, aperiodic and have an invariant distribution, $\lambda$ an arbitrary distribution, and suppose that $(X^t)_{t \ge 0}$ is Markov$(\lambda, P)$ \cite{chlithesis}, then
\begin{eqnarray}
& \Pr(X^t = j) \to p_{j}^\infty \ as\ t \to \infty\ \mbox{for all}\ j \in I & \\
& and & \nonumber \\
& p^{(t)}_{ij} \to p_{j}^\infty \ as\ t \to \infty\ \mbox{for all}\ i,j \in I. &
\end{eqnarray}
\end{theorem}

\vspace{5mm}

We can now view a system $S$ as a countable set of states $I$ with implicitly defined transitions between them, and at time $t$ the state of the system is the random variable $X^t$, with the key assumption that $(X^t)_{t \ge 0}$ is Markov$(\lambda,P)$ \addedold{\cite{chlithesis,msthesisNorris1997,msthesisCoxMiller1972}}.\\

\begin{definition}
The system $S$ is said to be stable when the distribution of its states converge to an \emph{equilibrium distribution},
\begin{equation}
\Pr(X^t = j) \to p_{j}^\infty \ as\ t \to \infty\ for\ all j\ \in I.
\end{equation}
\end{definition}

\vspace{5mm}

More intuitively, the system $S$, a stochastic process $X^0$, $X^1$, $X^2$, ... is \emph{stable} if the probability distribution of $X^t$ becomes independent of the time index $t$ for large $t$ \cite{chli2}. Most Markov chains with a finite state-space and positive transition probabilities are examples of controllable stable systems, because after an initialisation period they reach a stationary distribution \cite{chlithesis}.

A \acl{MAS} can be viewed as a system $S$, with the system state represented by a finite vector $\bx$, having dimensions large enough to represent the states of the agents in the system. The state vector will consist of one or more elements for each agent, and a number of elements to define general properties\addedold{\footnote{\addedold{These general properties are intended to represent properties that are external to the agents, and as such could include the coupling between the agents. However, we would expect such properties, as the coupling between the agents, to be stored within the agents themselves, and so be part of the elements defining the agents.}}} of the system state. \addedold{Hence there are many more states of the system (different state vectors) than there are agents.} 
We can then model agent \emph{death}, i.e. not being present in the system, by setting the vector elements for that agent to a fixed value that is shared by no other agent, called $d$ \cite{chlithesis}.

\section{Including Evolution}
\label{def}

\addedold{Having now introduced Chli-DeWilde stability, we will now briefly introduce Evolutionary Computing, sufficiently to allow for the derivation of our extensions to Chli-DeWilde stability to include \aclp{MAS} with Evolutionary Computing.} Evolution is the source of many diverse and creative solutions to problems in nature \cite{ec15, ec16}. However, it can also be useful as a problem-solving tool in artificial systems. Computer scientists and other theoreticians realised that the selection and mutation mechanisms that appear so effective in biological evolution could be abstracted to a computational algorithm \cite{ecpaper}. This Evolutionary Computing is now recognised as a sub-field of artificial intelligence (more particularly computational intelligence) that involves combinatorial optimisation problems \cite{ec17}.

\subsection{\added{Evolutionary Algorithms}}

Evolutionary algorithms are based upon several fundamental principles from biological evolution, including reproduction, mutation, recombination (crossover), natural selection, and survival of the fittest. As in biological populations, evolution occurs by the repeated application of the above operators \cite{back1996eat}. An evolutionary algorithm operates on the collection of individuals making up a population. An \emph{individual}, in the natural world, is an organism with an associated fitness \cite{lawrence1989hsd}. So, candidate solutions to an optimisation problem play the role of individuals in a population, and a cost (fitness) function determines the environment within which the solutions \emph{live}, analogous to the way the environment selects for the fittest individuals. The number of individuals varies between different implementations and may also vary during the use of an evolutionary algorithm. Each individual possesses some characteristics that are defined through its genotype, its genetic composition, which will be passed onto the descendants of that individual \cite{back1996eat}. Processes of mutation (small random changes) and crossover (generation of a new genotype by the combination of components from two individuals) may occur, resulting in new individuals with genotypes differing from the ancestors they will come to replace. These processes iterate, modifying the characteristics of the population \cite{back1996eat}. Which members of the population are kept, and used as parents for offspring, depends on the fitness (cost) function of the population. This enables improvement to occur \cite{back1996eat}, and corresponds to the fitness of an organism in the natural world \cite{lawrence1989hsd}. Recombination and mutation create the necessary diversity and thereby facilitate novelty, while selection acts as a force increasing quality. Changed pieces of information resulting from recombination and mutation are randomly chosen. Selection operators can be either deterministic, or stochastic. In the latter case, individuals with a higher fitness have a higher chance to be selected than individuals with a lower fitness \cite{back1996eat}.

\subsection{\added{Evolving Agent Populations}}

While the construction of \aclp{MAS} that make use of Evolutionary Computing differs \cite{smith1998fec, bionetics}, they universally include populations of agents evolving to provide desired functionality, i.e. \emph{evolving agent populations}. So, extending Chli-DeWilde stability to the class of \aclp{MAS} that make use of Evolutionary Computing requires understanding population dynamics and macro-states.

\subsubsection{\changed{Population Dynamics}}

\addedold{{\em Change of notation}. Section III has defined the main properties of Markov chains that we use to model Multi-Agent Systems. We have used the standard notation, as used in \cite{msthesisNorris1997} and many other textbooks on stochastic processes. This notion uses integers to denote states. These integers are easily used to label the rows and columns of the transition matrix. At the end of Section II, we made the step from Markov chain to multi-agent system. There are many more states than agents, and both ought to be labeled by integers. This turns out to be confusing, and therefore we make a {\em change of notation} here. We will now label the {\em agents} with integers $i$, and the states by vectors. Each agent $i$ is in a scalar state $\xi_i$. All $n$ agents together are in an $n$-dimensional vector state $\bxi$. (An agent can be described by multiple numbers, but we group these here into a single scalar $\xi_i$, just as multiple binary digits can be read as a single integer.)}

\addedold{The states of the agents are random variables. The actual values of the random variable $\xi_i$ will be denoted by numbers $X_i$ or $Y_i$. The actual values of the vector of random variables $\bxi$ will be denoted by vectors $\bx$ or $\by$. All the states can vary in time, and time will be denoted by the superscript $t$. The symbol for $t$ is the same as in Section II, but the symbols $i$ and $X$ have a new meaning. The symbol $i$ now denotes the agent, and $X_i$ is a number describing the state of the agent $i$.}

\addedold{The change of notation is essential because each agent is described by a random variable that has the Markov property. The multi-agent system is a network of interacting Markov chains. Without our change of notation, one would need multiple subscripts counting different things.}

A \acl{MAS} that makes use of Evolutionary Computing, an evolving agent population, is composed of $n$ agents, with each agent $i$ in a state $\xi_i^t$ at time $t$, where $i=1, 2, \ldots, n$. The states of the agents are \emph{random variables}, and so the state vector for the \acl{MAS} is a vector of random variables $\bxi^t$, with the time being discrete, $t=0, 1, \ldots$ . The interactions among the agents are noisy, and are given by the probability distributions
\be
\Pr(X_i | \by) = \Pr(\xi_i^{t+1} = X_i | \bxi^t = \by) , \quad \addedold{i=}\ra,
\eeq{eq1}
where $X_i$ is a value for the state of agent $i$, and $\by$ is a value for the state vector of the \acl{MAS}. The probabilities implement a Markov process \cite{suzuki}, with the noise caused by mutations. Furthermore, the agents are individually subjected to a \emph{selection pressure} from the environment of the system, which is applied equally to all the agents of the population. So, the probability distributions are statistically independent, and
\be
\Pr(\bx | \by) = \Pi_{i=1}^n \Pr(\xi_i^{t+1} = X_i | \bxi^t = \by).
\eeq{eq5}
If the occupation probability of state $\bx$ at time $t$ is denoted by $p_{\bx}^t$, then
\be
p_{\bx}^t = \sum_{\by} \Pr(\bx | \by) p_{\by}^{t-1}.
\eeq{eq5.1}
This is a discrete time equation used to calculate the evolution of the state occupation probabilities from $t=0$, while equation (\ref{eq5}) is the probability of moving from one state to another. The \acl{MAS} is self-stabilising if the limit distribution of the occupation probabilities exists and is non-uniform, i.e.
\be
p_\bx^\infty = lim_{t \rightarrow \infty} p_{\bx}^t
\eeq{eq2}
exists for all states $\bx$, and there are states $\bx$ and $\by$ such that
\be
p_\bx^\infty \neq p_\by^\infty.
\eeq{eq3}
These equations imply that some configurations of the system after an extended time will be more likely than others, because the likelihood of their occurrence no longer changes. Such a system is \emph{stable}, because the likelihood of states occurring no longer changes with time, and this is the definition of stability developed in \cite{chli2}. This is stability in the stochastic sense, because there is some indeterminacy in the future evolution described by the probability distributions. So, even with knowing the initial conditions, there are many directions in which the process might evolve, but still some paths will be more probable than others. Equation (\ref{eq2}) is the \emph{probabilistic equivalent} of an \emph{attractor}\footnote{An attractor is a set of states, invariant under the dynamics, towards which neighbouring states asymptotically approach during evolution \cite{weisstein2003cce}.} in a system with deterministic interactions, which we had to extend to a stochastic process because mutation is inherent in Evolutionary Computing.

While the number of agents in the Chli-DeWilde formalism can vary, we require it to vary according to the \emph{selection pressure} acting upon the evolving agent population. We must therefore formally define and extend the definition of \emph{dead} agents, by introducing a new state $d$ for each agent. If an agent is in this state, $\xi_i^t=d$, then it is \emph{dead} and does not affect the state of other agents in the population. If an agent $i$ has low fitness then that agent will likely die, because
\be
\Pr(d | \by) = \Pr(\xi_i^{t+1} = d | \bxi^t = \by)
\eeq{eq4}
will be high for all $\by$. Conversely, if an agent has high fitness, then it will likely replicate, becoming a similarly successful agent (mutant), or crossover might occur changing the state of the successful agent and another agent.

\subsubsection{\changed{Population Macro-States}}

\changedold{The state of the system, an evolving agent population, $S$ is determined by the collection of agents of which it consists at a specific time $t$, which potentially changes as the time increases, $t+1$.} \addedold{This collection of agents will have varying fitness values, and so the one with the highest fitness at the current time $t$ is the \emph{current maximum fitness individual}. For example, an evolving agent population with individuals ranging in fitness between 36.2\% \changed{and} 45.8\%, the \emph{current maximum fitness individual (agent)} is the one with a fitness of 45.8\%. So,} \changedold{w}e can define a macro-state $M$ as a set of states (evolving agent populations) with a common property, here possessing at least one copy of the \emph{current maximum fitness individual}. Therefore, by its definition, each macro-state $M$ must also have a \emph{maximal state} composed entirely of copies of the \emph{current maximum fitness individual}. \addedold{If the population size is not fixed (not in nature, can be in evolutionary computing), the state space of the evolving agent population is infinite, but in practise would be bounded by resource availability. So, there is also an infinite number of configurations for an evolving agent population that has the same \emph{current maximum fitness individual}.}

\addedold{So, the state-space $I$ of the system (evolving agent population) $S$ can be grouped to a set macro-states $\{M\}$. For one macro-state, which we will call the \emph{maximum macro-state} $M_{max}$, the \emph{current maximum fitness individual} will be the \emph{global maximum fitness individual}, which is the \emph{optimal solution} (\emph{fittest} individual) that the evolutionary computing process can reach through the evolving agent population (system) $S$. \addedold{For example, an evolving agent population at its \emph{maximum macro-state} $M_{max}$, with individuals ranging in fitness between 88.8\% \changed{and} 96.8\%, the \emph{global maximum fitness individual (agent)} is the one with a fitness of 96.8\% and there will be no \emph{fitter} agent.} Also, we can therefore refer to all other macro-states of the system $S$ as \emph{sub-optimal} macro-states, as there can be only one \emph{maximum macro-state} $M_{max}$.}

\tfigure{width=3.5in}{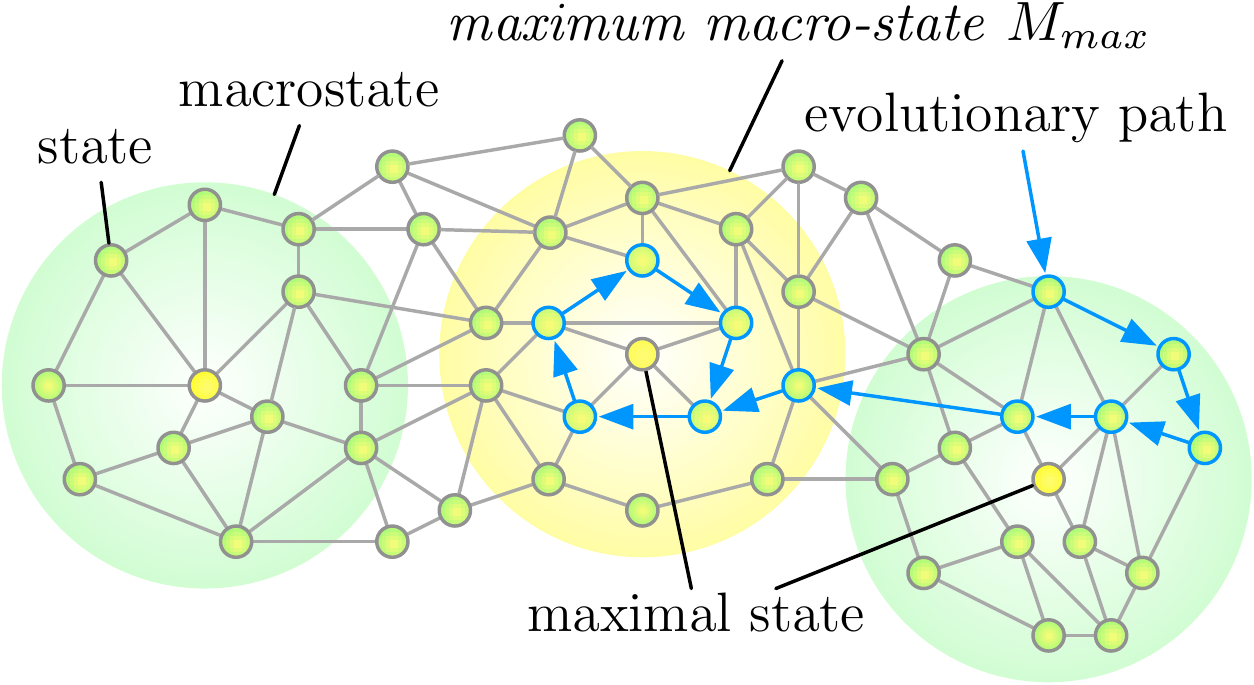}{graffle}{State-Space of an Evolving Agent Population}{A \getCap{statesCap} is shown, with \getCap{capStates3} \getCap{capStates}.}{-6mm}{}{}{}

We can consider the \emph{macro-states} of an evolving agent population visually through the representation of the state-space $I$ of the system $S$ shown in Figure \ref{fig1}, which includes a \setCap{possible evolutionary path through the state-space}{statesCap}. Traversal through the state-space $I$ is directed by \setCap{the \emph{selection pressure} of the evolutionary process}{capStates3} acting upon the population $S$, \setCap{driving it towards the \emph{maximal state} of the maximum macro-state $M_{max}$}{capStates}, consisting entirely of copies of the \emph{optimal solution}. It is the equilibrium state that the system $S$ is forever falling towards without ever quite reaching, because of the noise (mutation) within the system. Yet, the maximum macro-state $M_{max}$, in which this \emph{maximal state} is located, will be reached, provided the system does not get trapped at local optima, i.e. the probability of being in the maximum macro-state $M_{max}$ at infinite time is one, $p^{\infty}_{\bone}=1$.

Furthermore, we can define quantitatively the probability distribution of the macro-states that the system occupies at infinite time. For a stable system, as defined by equation (\ref{eq3}), the \emph{degree of instability}, $\delta$, is the entropy of its probability distribution at infinite time,
\be
\delta = H(p^\infty) = -\sum\limits_{\bx}p_{\bx}^{\infty}log_{N}(p_{\bx}^{\infty}),
\eeq{eq6}
where $N$ is the number of possible states, and taking $log$ to the base $N$ normalises the \emph{degree of instability}. The \emph{degree of instability} will range between zero (inclusive) and one (exclusive), because a maximum instability of one would only occur in the theoretical extreme scenario of a \emph{non-discriminating selection pressure} \cite{kimura:ntm}.

\section{Simulation and Results}
\label{simandres}

\subsection{\added{Evolutionary Dynamics}}

\changedtwo{A simulated agent population was evolved relative to an artificial \emph{selection pressure} created by a \emph{fitness function} generated from a user request $R$. An individual (agent) of the population consisted of a set of attributes, ${a_1, a_2, ...}$, and a user request consisted of a set of required attributes, ${r_1, r_2, ...}$. The \emph{fitness function} for evaluating an individual agent $A$, relative to a user request $R$, was
\begin{equation}
fitness(A,R) = \frac{1}{1 + \sum_{r \in R}{|r-a|}},
\label{ff}
\end{equation}
where $a$ is an attribute of the agent $A$ such that the difference to the required attribute $r$ of $R$ was minimised. \addedold{The abstract agent descriptions was based on existing and emerging technologies for \emph{semantically capable} \aclp{SOA} \cite{SOAsemantic}, such as the \acs{OWL-S} semantic markup for web services \cite{martin2004bsw}. We simulated an agent's \emph{semantic description} \setCap{with an abstract representation consisting of a set of}{as3} attributes, to simulate the properties of a \emph{semantic description}. Each \setCap{attribute representing a \emph{property} of the \emph{semantic description}, ranging between one and a hundred.}{agentSemantic2} \setCap{Each simulated agent was initialised with a semantic description}{as4} of between three and six attributes, which would then evolve in number and content.}}

\changedtwo{Equation \ref{ff} was used to assign \emph{fitness} values between 0.0 and 1.0 to each individual of the current generation of the population, directly affecting their ability to replicate into the next generation. The Evolutionary Computing process was encoded with a low mutation rate, a fixed selection pressure and a non-trapping fitness function (i.e. did not get trapped at local optima\addedold{\footnote{\addedold{These constraints can be considered in abstract using the metaphor of the \emph{fitness landscape}, in which individuals are represented as solutions to the problem of survival and reproduction \cite{wright1932}. All possible solutions are distributed in a space whose dimensions are the possible properties of individuals. An additional dimension, height, indicates the relative fitness (in terms of survival and reproduction) of each solution. The fitness landscape is envisaged as a rugged, multidimensional landscape of hills, mountains, and valleys, because individuals with certain sets of properties are \emph{fitter} than others \cite{wright1932}. Here the ruggedness of the fitness landscape is not so severe, relative to the population diversity (population size and mutation rate), to prevent the evolving population from progressing to the global optima of the fitness landscape.}}}).}

\changedtwo{The type of selection used was fitness-proportional and non-elitist, with fitness-proportional meaning that the \emph{fitter} the individual the higher its probability of surviving to the next generation. Non-elitist means that the best individual from one generation was not guaranteed to survive to the next generation; it had a high probability of surviving into the next generation, but it was not guaranteed as it might have been mutated \cite{eiben2003iec}. \addedold{\changedtwo{These} initial \changed{parameters} where chosen to focus on studying our extended Chli-DeWilde stability, rather than the evolutionary computing process itself.}}

\changedtwo{\emph{Crossover} (recombination) was then applied to a randomly chosen 10\% of the surviving population. \emph{Mutations} were then applied to a randomly chosen 10\% of the surviving population; \emph{point mutations} were randomly located, consisting of \emph{insertions} (an attribute was inserted into an agent), \emph{replacements} (an attribute was replaced in an agent), and \emph{deletions} (an attribute was deleted from an agent). \addedold{Rates \changedtwo{of} 10\% where chosen, because they would provide the necessary behaviour to a sufficient degree.} A dynamic population size was used\addedold{, with an initial population size of 300,} to ensure exploration of the available agent attribute combination space, which increased with the average size of the population's agents, \addedold{because fixed population sizes can sometimes fail to sufficient search available combination spaces.}}

The issue of bloat \cite{ec25} was controlled by augmenting the \emph{fitness function} with a \emph{parsimony pressure} \cite{soule1998ecg}, which biased the search to smaller agents, evaluating larger than average agents with a reduced \emph{fitness}, and thereby providing a dynamic control limit which adapted to the average size of the individuals of the evolving agent population.

\added{We first plotted the fitness of the evolutionary process for a typical evolving agent population to elucidate its inherent \emph{evolutionary dynamics}. The graph in Figure \ref{fig2} \setCap{shows both the maximum and average fitness increasing over the generations of a typical evolving agent population, and as expected the \emph{average fitness} remains below the \emph{maximum fitness} because of variation in the evolving agent population \cite{goldberg}}{evoGraph}, showing that the inherent dynamism of evolutionary processes applies to evolving agent populations.}

\tfigure{width=3.5in}{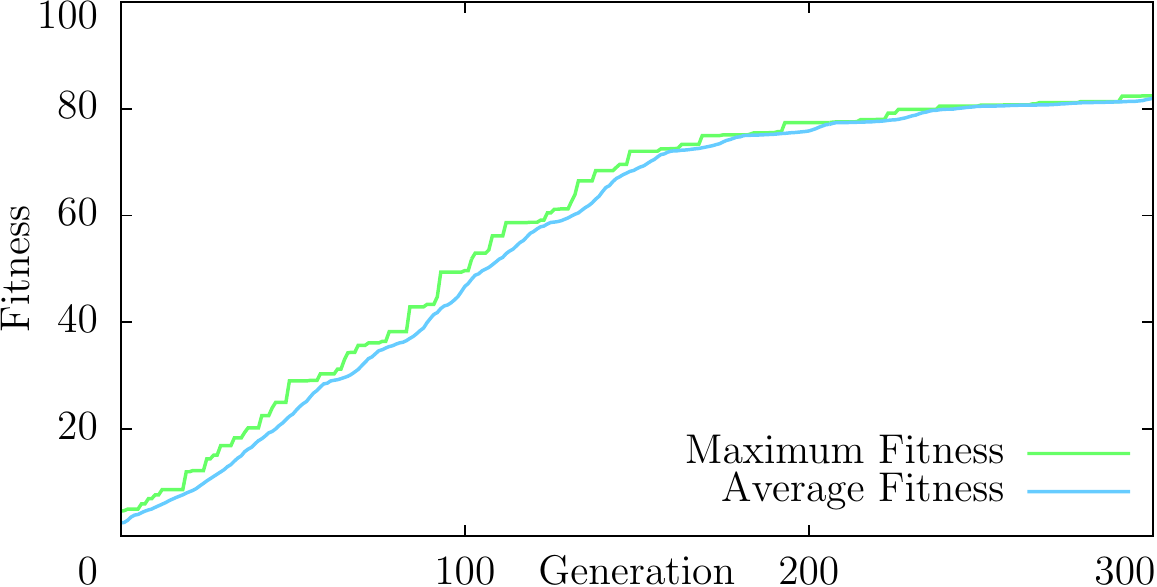}{graph}{\added{Graph of Evolutionary Dynamics}}{\added{This \getCap{evoGraph}.}}{-5mm}{!t}{}{-5mm}

\subsection{\addedold{Stability}}

\subsubsection{\addedold{Initial Parameters}}
An evolving agent population was called stable if the distribution of the limit probabilities existed and was non-uniform, as defined by equations (\ref{eq2}) and (\ref{eq3}). The simplest case was a typical evolving agent population with a global optimum, which was stable if there were at least two macro-states with different limit occupation probabilities. So, we considered the maximum macro-state $M_{max}$ and \changedold{one of the} \emph{sub-optimal macro-state\changedold{s}}\addedold{,} $M_{half}$. Where the states of the macro-state $M_{max}$ each possessed at least one individual with global maximum fitness,
\begin{equation*}
p_{\bone}^\infty = lim_{t \rightarrow \infty} p_{\bone}^{(t)} = 1,
\end{equation*}
while the states of the macro-state $M_{half}$ each possessed at least one individual with fitness equal to \emph{half} of the global maximum fitness,
\begin{equation*}
p_{\bfif}^\infty = lim_{t \rightarrow \infty} p_{\bfif}^{(t)} = 0,
\end{equation*}
thereby fulfilling the requirements of equations (\ref{eq2}) and (\ref{eq3}). A value of $t=1000$ was chosen to represent $t=\infty$ experimentally, because the simulation has often been observed to reach the maximum macro-state $M_{max}$ within 500 generations. Therefore, the probability of the system $S$ being in the maximum macro-state $M_{max}$ at the thousandth generation is expected to be one, $p^{1000}_{\bone} = 1$. Furthermore, the probability of the system being in the sub-optimal macro-state $M_{half}$ at the thousandth generation is expected to be zero, $p^{1000}_{\bfif} = 0$.

\subsubsection{\addedold{Predictions}}
The sub-optimal macro-state $M_{half}$, having a lower fitness, is predicted to be seen earlier in the evolutionary process before disappearing as higher fitness macro-states are reached. The system $S$ will take longer to reach the maximum macro-state $M_{max}$, but once it does will likely remain, leaving only briefly depending on the strength of the mutation rate, as the \emph{selection pressure} was non-elitist.

\tfigure{width=3.5in}{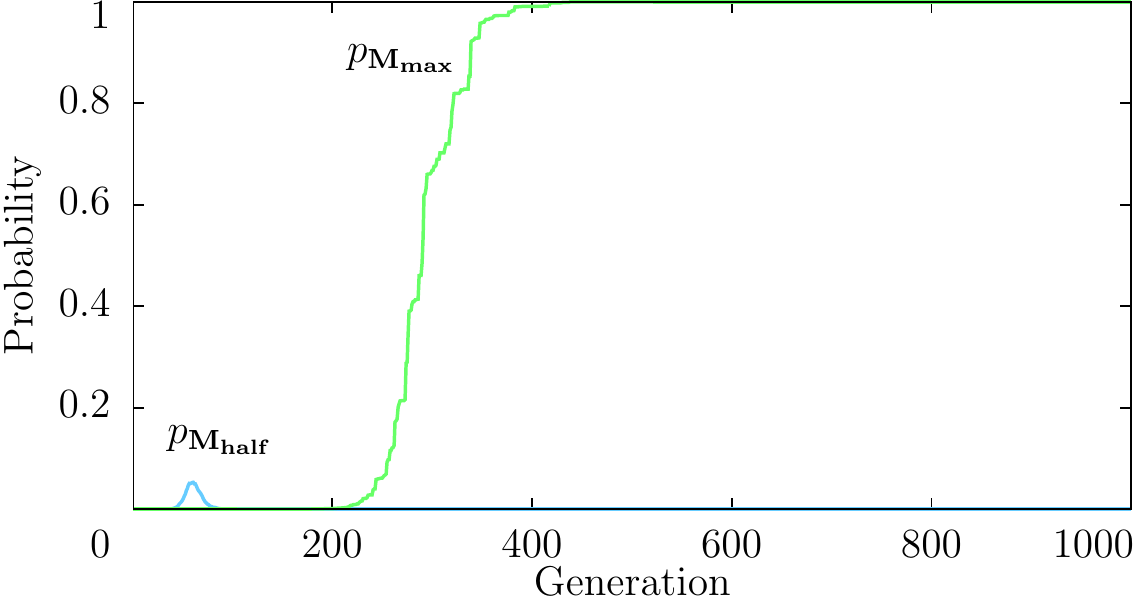}{graph}{Graph of the Probabilities of the Macro-States $M_{max}$ and $M_{half}$ at each Generation}{The system $S$, a typical evolving agent population, was \getCap{graphCap}.}{-6mm}{!b}{}{}

\subsubsection{\addedold{Results}}
Figure \ref{fig3} shows, for a typical evolving agent population, a graph of the probability as defined by equation (\ref{eq5.1}) of the maximum macro-state $M_{max}$ and the sub-optimal macro-state $M_{half}$ at each generation, averaged from ten thousand simulation runs for statistical significance. The behaviour of the simulated system $S$ was as expected, being \setCap{in the maximum macro-state $M_{max}$ only after generation 178 and always after generation 482. It was also observed being in the sub-optimal macro-state $M_{half}$ only between generations 37 and 113}{graphCap}, with a maximum probability of 0.053 at generation 61. This is because the evolutionary path (state transitions) could avoid visiting the macro-state.

\subsubsection{\addedold{Conclusions}}
As expected the probability of being in the maximum macro-state $M_{max}$ at the thousandth generation was one, $p^{1000}_{\bone} = 1$, and so the probability of being in any other macro-state, including the sub-optimal macro-state $M_{half}$, at the thousandth generation was zero, $p^{1000}_{\bfif} = 0$. \addedold{We can therefore conclude that our extended Chli-DeWilde stability accurately models the stability over time of evolving agent populations (\aclp{MAS} with Evolutionary Computing).}

\subsection{Visualisation}

\subsubsection{\addedold{Results}}
A visualisation for the state of a typical evolving agent population\addedold{, from the experiment of previous subsection,} at the thousandth generation is shown in Figure \ref{fig4}, with each line representing an agent and each shade representing an agent attribute, with the identical agents grouped for clarity. It shows that the evolving agent population reached the maximum macro-state $M_{max}$ and remained there, but as expected never reached its \emph{maximal state}, where all the agents are identical and have maximum fitness, which is indicated by the lack of total uniformity in the visualisation.

\tfigure{width=3.5in}{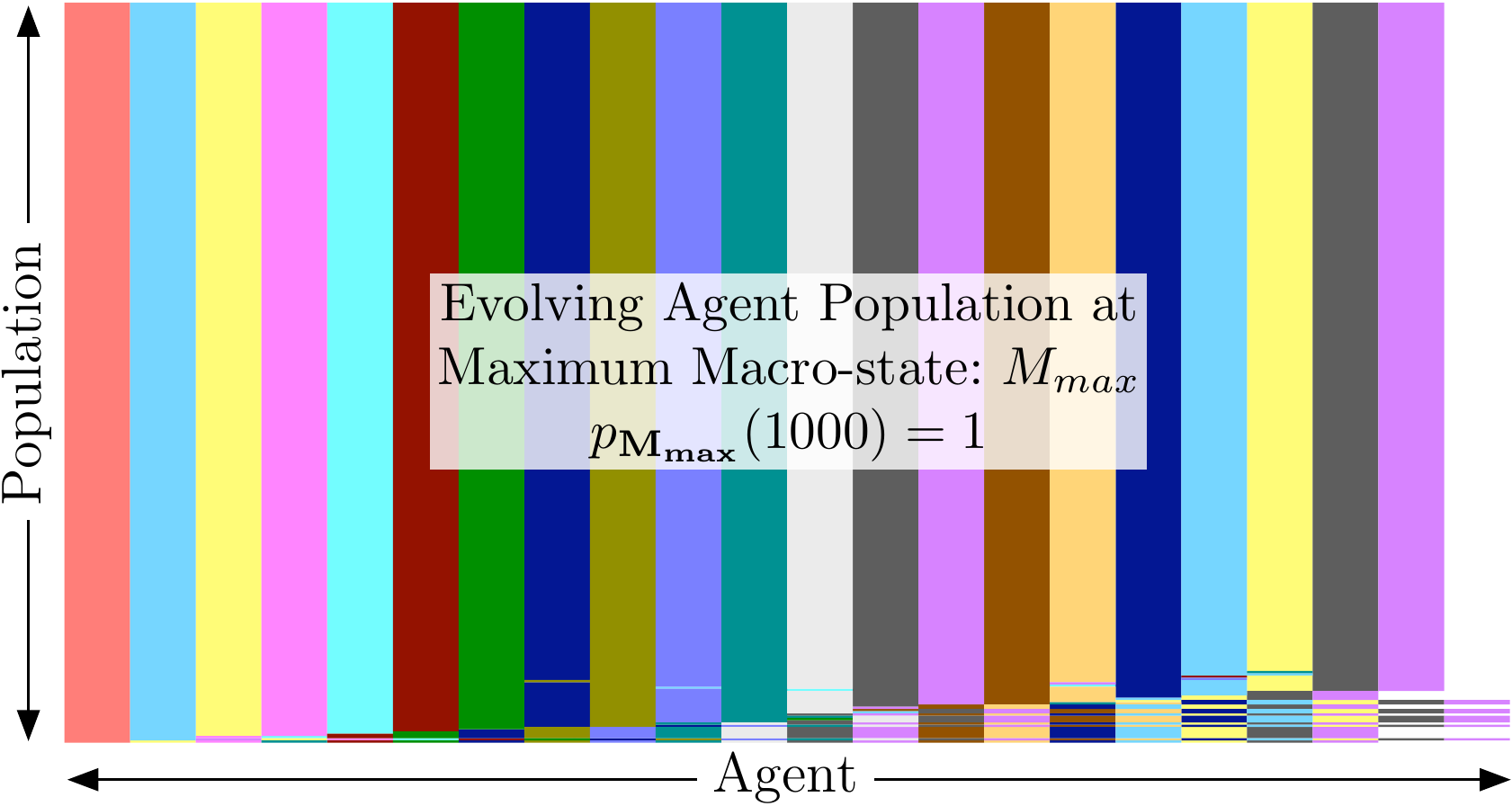}{graffle}{Visualisation of an Evolving Agent Population at the 1000th Generation}{The population consists of \changedold{323} agents, with each line representing an agent, \changedold{and each} shade representing an agent attribute. \addedold{So, we see a population consisting of multiple agents, many of which are identical, having the same maximum global fitness.} The identical agents were grouped for clarity, and as expected the system $S$ reached the {maximum macro-state} $M_{max}$.}{-7mm}{!h}{2.5mm}{2.5mm}

\subsubsection{\addedold{Conclusions}}
\addedold{The result} was \addedold{as} expected\addedold{, a lack of total uniformity in the visualisation,} because of the mutation (noise) within the evolutionary process, which is necessary to create the opportunity to find fitter (better) sequences and potentially avoid getting trapped at any local optima that may be present. \addedold{We can therefore conclude that the macro-state interpretation of our extended Chli-DeWilde stability accurately models the state-space of evolving agent populations (\aclp{MAS} with Evolutionary Computing).}

\subsection{Degree of Instability}

\subsubsection{\addedold{Results}}
Given that our simulated evolving agent population is stable as defined by equations (\ref{eq2}) and (\ref{eq3}), we can determine the \emph{degree of instability} as defined by equation (\ref{eq6}). So, calculated from its limit probabilities, the \emph{degree of instability} was
\begin{eqnarray}
\delta = H(p^{1000}) &=& -\sum\limits_{\bx}p^{1000}_\bx log_{N}(p^{1000}_\bx) \nonumber \\
 &=& -1log_{N}(1) \nonumber \\
 &=& 0, \nonumber
\label{result2}
\end{eqnarray}
where $t=1000$ was an effective estimate for $t=\infty$. This result was expected because the maximum macro-state $M_{max}$ at the thousandth generation was one, $p^{1000}_{\bone} = 1$, and so the probability of being in any other macro-state at the thousandth generation was zero. 

\subsubsection{\addedold{Conclusions}}
The system therefore showed no instability, as there is no entropy in the occupied macro-states at infinite time. \addedold{We can therefore conclude that the \emph{degree of instability} of our extended Chli-DeWilde stability can provide a macroscopic value to characterise the \emph{level of stability} of evolving agent populations (\aclp{MAS} with Evolutionary Computing).}

\subsection{Stability Analysis}

\subsubsection{\addedold{Initial Parameters}}
We then performed a \emph{stability analysis} (similar to a \emph{sensitivity analysis} \cite{cacuci2003sau}) of a typical evolving agent population, by varying its key parameters while measuring its stability. We varied the mutation and crossover rates from 0\% to 100\% in 10\% increments \addedold{to provide a sufficient density of measurements to identify any trends that might be present}, calculating the \emph{degree of instability}, $\delta$ from (\ref{eq6}), at the thousandth generation. These \emph{degree of instability} values were averaged over 10 000 simulation runs \addedold{to ensure statistical significance}, and graphed against the mutation and crossover rates in Figure \ref{fig5}. 

\tfigure{width=3.5in}{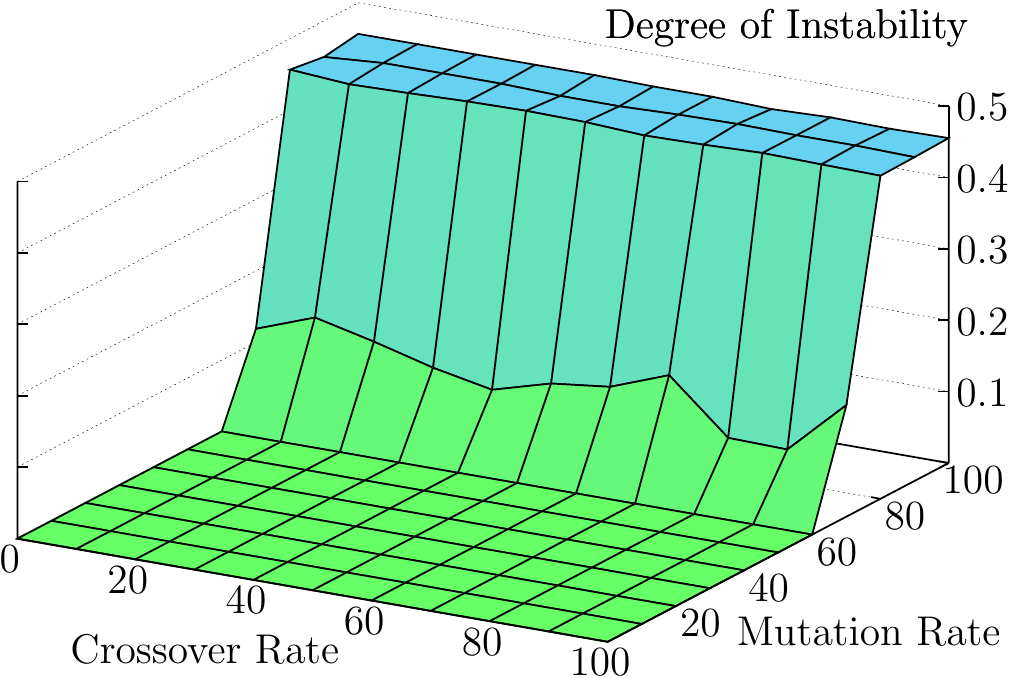}{graph}{Graph of Stability with Different Mutation and Crossover Rates}{\getCap{aScap}.}{-6mm}{!b}{}{}

\subsubsection{\addedold{Results}}
It shows that the crossover rate had little effect on the stability of the evolving agent population, whereas the mutation rate did significantly affect stability. \setCap{With the mutation rate under or equal to 60\%, the evolving agent population showed no instability, with $\delta$ values equal to zero as the system $S$ was always in the same macro-state $M$ at infinite time, independent of the crossover rate. With the mutation rate above 60\% the instability increased significantly}{aScap}, with the system being in one of several different macro-states at infinite time; with a mutation rate of 70\% the system was still very stable, having low $\delta$ values ranging between 0.08 and 0.16, but once the mutation rate was 80\% or greater the system became quite unstable, shown by high $\delta$ values nearing 0.5.

\subsubsection{\addedold{Conclusions}}
As one would have expected, an extremely high mutation rate had a destabilising \changed{affect} on the stability of evolving agent populations. \addedold{Also, as expected} \changedold{t}he crossover rate had only a minimal effect, because variation from crossover was limited once the population matured, consisting of agents identical or very similar to one another. \changed{It should also be noted that the stability of the system is different to its performance at optimising, because while showing no instability with mutation rates below 60\% (inclusive), it reached the maximum macro-state $M_{max}$ only with a mutation rate of 10\% or above, while at 0\% it was stable at sub-optimal macro-state\added{s between $M_{tenth}$ (inclusive) and $M_{twentieth}$ (inclusive), i.e. with at least one individual with a fitness between \emph{twentieth} (inclusive) to a \emph{tenth} (inclusive) of the global maximum fitness, as indicated by all of the 0\% Mutation Rate experiments have degree of instability, $\delta$, values of zero in Figure \ref{fig5}. This is because there was no mutation (mutate rate = 0\%), and so the evolving agent populations always remained near the sub-optimal macro-state to which they where initialised (seeded), with any crossover rate having little effect.}}

\addedold{We can therefore conclude that the \emph{degree of instability} of our extended Chli-DeWilde stability can used of perform stability analyses (similar to a \emph{sensitivity analyses} \cite{cacuci2003sau}) of evolving agent populations (\aclp{MAS} with Evolutionary Computing).}

\section{Conclusions}

Our extension of Chli-DeWilde stability was developed to provide a greater understanding of stability in \aclp{MAS} that make use of Evolutionary Computing, i.e. evolving agent populations. We then built upon this to construct an entropy-based definition for the \emph{degree of instability}, which provides information about the level of stability, applicable to \aclp{MAS} with or without Evolutionary Computing. Furthermore, it can be used to perform a \emph{stability analysis}, similar to a \emph{sensitivity analysis}, of \aclp{MAS}.

Collectively, the experimental results confirm that Chli-DeWilde stability has been successfully extended to evolving agent populations, while our definition for the \emph{degree of instability} provides a macroscopic value to characterise the \emph{level of stability}. These findings also support the proposition that Chli-DeWilde stability can be widely applied to different classes of \aclp{MAS}. So, our extended Chli-DeWilde stability is a useful tool for analysing \aclp{MAS}, with or without Evolutionary Computing, providing an effective understanding and quantification to help better understand the stability of such systems.

Overall, an insight has been achieved into the stability of \aclp{MAS} that make use of Evolutionary Computing, which is a first step in being able to control such systems. For example, say one wanted to avoid a number of \emph{bad} states. If the probability of being in those states kept changing with time, it would be difficult to devise a strategy to avoid these states. However, if the probability converges in time, while one could not guarantee to avoid those states, one could at least calculate the expected damage, i.e. the probability of being in a state times by the penalty for being in it, summed over all the states one wishes to avoid. Stochastic control of multi-agent systems will be the subject of further work. 

\changed{\addedold{Our future work will also consider include more experimental scenarios to further consolidate the conclusions, with a range of different \emph{fitness landscapes} \cite{wright1932}, including flat ones (from the neutral theory of molecular evolution \cite{kimura:ntm}) and ones with multiple global optima.}}

\section{Acknowledgments}

The authors would like to thank the following for encouragement and suggestions; Dr Paolo Dini of the London School of Economics and Political Science, Dr Maria Chli of Aston University, and Dr Jon Rowe of the University of Birmingham. This work was supported by the EU-funded OPAALS Network of Excellence (NoE), Contract No. FP6/IST-034824.

\bibliographystyle{plain}
\nocite{ec24, mspaperMarwala2001}
\bibliography{references,myRefs}

\begin{thebibliography}{10}

\bibitem{angeli2006stability}
D.~Angeli and P.A. Bliman.
\newblock Stability of leaderless discrete-time multi-agent systems.
\newblock {\em Mathematics of Control, Signals, and Systems (MCSS)},
  18(4):293--322, 2006.

\bibitem{back1996eat}
T.~B{\"a}ck.
\newblock {\em Evolutionary Algorithms in Theory and Practice: Evolution
  Strategies, Evolutionary Programming, Genetic Algorithms}.
\newblock Oxford University Press, 1996.

\bibitem{ec17}
T~Baeck, D~Fogel, and Z~Michalewicz, editors.
\newblock {\em Handbook of Evolutionary Computation}.
\newblock CRC Press, 1997.

\bibitem{mspaper9Balakrishnan1997}
H.~Balakrishnan, M.~Stemm, S.~Seshan, and R.~Katz.
\newblock Analyzing stability in wide-area network performance.
\newblock In Scott Leutenegger, editor, {\em International Conference on
  Measurement and Modeling of Computer Systems}, pages 2--12. ACM Press, 1997.

\bibitem{thesis}
G~Briscoe.
\newblock {\em Digital Ecosystems}.
\newblock PhD thesis, Imperial College London, 2009.

\bibitem{caes}
G~Briscoe.
\newblock Complex adaptive digital ecosystems.
\newblock In {\em ACM Management of Emergent Digital Ecosystems Conference},
  2010.

\bibitem{bionetics}
G~Briscoe and P~{De Wilde}.
\newblock Digital {E}cosystems: Evolving service-oriented architectures.
\newblock In {\em IEEE Bio Inspired Models of Network, Information and
  Computing Systems Conference}, 2006.

\bibitem{acmMedes}
G~Briscoe and P~De~Wilde.
\newblock Computing of applied digital ecosystems.
\newblock In {\em ACM Management of Emergent Digital Ecosystems Conference},
  2009.

\bibitem{agentStability}
G~Briscoe and P~{De Wilde}.
\newblock Digital {E}cosystems: Stability of evolving agent populations.
\newblock In {\em ACM Management of Emergent Digital Ecosystems Conference},
  2009.

\bibitem{epi}
G~Briscoe and P~{De Wilde}.
\newblock The computing of digital ecosystems.
\newblock {\em International Journal of Organizational and Collective
  Intelligence}, 1(4), 2010.

\bibitem{dbebkpub}
G~Briscoe and S~Sadedin.
\newblock Natural science paradigms.
\newblock In {\em Digital {B}usiness {E}cosystems}, pages 48--55. European
  {C}ommission, 2007.

\bibitem{de07oz}
G~Briscoe, S~Sadedin, and G~Paperin.
\newblock Biology of applied digital ecosystems.
\newblock In {\em IEEE Digital Ecosystems and Technologies Conference}, pages
  458--463, 2007.

\bibitem{cacuci2003sau}
Dan Cacuci, Mihaela Ionescu-Bujor, and Ionel Navon.
\newblock {\em Sensitivity and Uncertainty Analysis}.
\newblock CRC Press, 2003.

\bibitem{chlithesis}
M~Chli.
\newblock {\em Convergence and Interactivity of Multi-Agent Systems}.
\newblock PhD thesis, Imperial College London, 2006.

\bibitem{chli2}
M~Chli, P~De~Wilde, J~Goossenaerts, V~Abramov, N~Szirbik, L~Correia, P~Mariano,
  and R~Ribeiro.
\newblock Stability of multi-agent systems.
\newblock In E~Santos~Jr and P~Willett, editors, {\em International Conference
  on Systems, Man, and Cybernetics}, pages 551--556. IEEE Press, 2003.

\bibitem{msthesisCoxMiller1972}
D.~Cox and H.~Miller.
\newblock {\em The Theory of Stochastic Processes}.
\newblock CRC Press, 1977.

\bibitem{ec15}
C~Darwin.
\newblock {\em On the Origin of Species by Means of Natural Selection}.
\newblock John Murray, 1859.

\bibitem{ec24}
K~De~Jong, D~Fogel, and H~Schwefel.
\newblock A history of evolutionary computation.
\newblock In Baeck et~al. \cite{ec17}, pages 1--12.

\bibitem{mspaperDeWilde1999a}
P.~De~Wilde, H.~Nwana, and L.~Lee.
\newblock Stability, fairness and scalability of multi-agent systems.
\newblock {\em International Journal of Knowledge-Based Intelligent Engineering
  Systems}, 3:84--91, 1999.

\bibitem{eibenAarts}
A~Eiben, E~Aarts, and K~Van~Hee.
\newblock Global convergence of genetic algorithms: A {M}arkov chain analysis.
\newblock In H~Schwefel and R~Manner, editors, {\em Parallel Problem Solving
  from Nature}, pages 4--12. Springer, 1991.

\bibitem{eiben2003iec}
A.~Eiben and J.~Smith.
\newblock {\em Introduction to Evolutionary Computing}.
\newblock Springer, 2003.

\bibitem{ec16}
D~Futuyma.
\newblock {\em Evolutionary Biology}.
\newblock Sinauer Associates, 1998.

\bibitem{goldberg}
D~Goldberg.
\newblock {\em Genetic algorithms in search, optimization, and machine
  learning}.
\newblock Addison-Wesley, 1989.

\bibitem{goldberg2}
D~Goldberg and P~Segrest.
\newblock Finite {M}arkov chain analysis of genetic algorithms.
\newblock In John Grefenstette, editor, {\em International Conference on
  Genetic Algorithms and their application}, pages 1--8. Lawrence Erlbaum
  Associates, 1987.

\bibitem{kimura:ntm}
M.~Kimura.
\newblock {\em The neutral theory of molecular evolution.}
\newblock Cambridge University Press, 1983.

\bibitem{ec25}
J~Koza.
\newblock {\em Genetic Programming: On the Programming of Computers by Means of
  Natural Selection}.
\newblock MIT Press, 1992.

\bibitem{lawrence1989hsd}
E.~Lawrence.
\newblock {\em Henderson's dictionary of biological terms}.
\newblock Pearson Education, 2005.

\bibitem{mspaperLee1998}
LC~Lee, HS~Nwana, DT~Ndumu, and P.~De~Wilde.
\newblock The stability, scalability and performance of multi-agent systems.
\newblock {\em BT Technology Journal}, 16:94--103, 1998.

\bibitem{mabu2007gbe}
S.~Mabu, K.~Hirasawa, and J.~Hu.
\newblock A graph-based evolutionary algorithm: Genetic network programming
  ({GNP}) and its extension using reinforcement learning.
\newblock {\em Evolutionary Computation}, 15:369--398, 2007.

\bibitem{ecpaper}
P~Marrow.
\newblock Nature-inspired computing technology and applications.
\newblock {\em BT Technology Journal}, 18:13--23, 2000.

\bibitem{martin2004bsw}
D.~Martin, M.~Paolucci, S.~McIlraith, M.~Burstein, D.~McDermott, D.~McGuinness,
  B.~Parsia, T.~Payne, M.~Sabou, M.~Solanki, Naveen Srinivasan, and Katia
  Sycara.
\newblock Bringing semantics to web services: The {OWL-S} approach.
\newblock In Jorge Cardoso and Amit Sheth, editors, {\em Semantic Web Services
  and Web Process Composition}, pages 6--9. Springer, 2004.

\bibitem{mspaperMarwala2001}
T.~Marwala, P.~De~Wilde, L.~Correia, P.~Mariano, R.~Ribeiro, V.~Abramov,
  N.~Szirbik, and J.~Goossenaerts.
\newblock Scalability and optimisation of a committee of agents using genetic
  algorithm.
\newblock In D~Campbell and C~Fyfe, editors, {\em International ICSC Symposium
  Soft Computing and Intelligent Systems For Industry}. ICSC-NAISO Academic
  Press, 2001.

\bibitem{mohanarajah2008formation}
G.~Mohanarajah and T.~Hayakawa.
\newblock Formation stability of multi-agent systems with limited information.
\newblock In {\em American Control Conference, 2008}, pages 704--709, 2008.

\bibitem{moreau2005sms}
L.~Moreau.
\newblock Stability of multiagent systems with time-dependent communication
  links.
\newblock {\em IEEE Transactions on Automatic Control}, 50:169--182, 2005.

\bibitem{dbebook}
F~Nachira, A~Nicolai, P~Dini, M~Le~Louarn, and L~Rivera~Le{\'o}n, editors.
\newblock {\em Digital {B}usiness {E}cosystems}.
\newblock European {C}ommission, 2007.

\bibitem{nix}
A.~Nix and M.~Vose.
\newblock Modeling genetic algorithms with {M}arkov chains.
\newblock {\em Annals of Mathematics and Artificial Intelligence}, 5:79--88,
  1992.

\bibitem{msthesisNorris1997}
J.~Norris.
\newblock {\em {M}arkov Chains}.
\newblock Cambridge University Press, 1997.

\bibitem{masOverviewPaper}
H~Nwana.
\newblock Software agents: An overview.
\newblock {\em Knowledge Engineering Review}, 11:205--244, 1996.

\bibitem{nwana1996software}
H.S. Nwana.
\newblock Software agents: An overview.
\newblock {\em Knowledge Engineering Review}, 11(3):205--244, 1996.

\bibitem{olfatisaber2007cac}
R.~Olfati-Saber, J.~Fax, and R.~Murray.
\newblock Consensus and cooperation in networked multi-agent systems.
\newblock {\em Proceedings of the IEEE}, 95:215--233, 2007.

\bibitem{SOAsemantic}
P~Rajasekaran, J~Miller, K~Verma, and A~Sheth.
\newblock Enhancing web services description and discovery to facilitate
  composition.
\newblock In Jorge Cardoso and Amit Sheth, editors, {\em Semantic Web Services
  and Web Process Composition}, pages 55--68. Springer, 2004.

\bibitem{rogers2007effects}
A.~Rogers, E.~David, N.R. Jennings, and J.~Schiff.
\newblock {The effects of proxy bidding and minimum bid increments within eBay
  auctions}.
\newblock {\em ACM Transactions on the Web (TWEB)}, 1(2):9, 2007.

\bibitem{rudolph}
G~Rudolph.
\newblock Convergence analysis of canonical genetic algorithms.
\newblock {\em IEEE Transactions on Neural Networks}, 5:96--101, 1994.

\bibitem{rudolph1998fmc}
G.~Rudolph.
\newblock Finite {M}arkov chain results in evolutionary computation: A tour
  d'horizon.
\newblock {\em Fundamenta Informaticae}, 35:67--89, 1998.

\bibitem{schurr2005future}
N.~Schurr, J.~Marecki, M.~Tambe, P.~Scerri, N.~Kasinadhuni, and J.~Lewis.
\newblock {The future of disaster response: Humans working with multiagent
  teams using DEFACTO}.
\newblock In {\em AAAI Spring Symposium on Homeland Security}, 2005.

\bibitem{smith2000eec}
R.~Smith, C.~Bonacina, P.~Kearney, and W.~Merlat.
\newblock Embodiment of evolutionary computation in general agents.
\newblock {\em Evolutionary Computation}, 8:475--493, 2000.

\bibitem{smith1998fec}
R.~Smith and N.~Taylor.
\newblock A framework for evolutionary computation in agent-based systems.
\newblock In Janice Glasgow, editor, {\em International Conference on
  Intelligent Systems}, pages 221--224. AAAI Press, 1998.

\bibitem{soule1998ecg}
T.~Soule and J.~Foster.
\newblock Effects of code growth and parsimony pressure on populations in
  genetic programming.
\newblock {\em Evolutionary Computation}, 6:293--309, 1998.

\bibitem{abcdbe}
J~Stanley and G~Briscoe.
\newblock The abc of digital business ecosystems.
\newblock {\em Communications Law - Journal of Computer, Media and
  Telecommunications Law}, 15(1), 2010.

\bibitem{sun2004simulating}
R.~Sun and I.~Naveh.
\newblock {Simulating organizational decision-making using a cognitively
  realistic agent model}.
\newblock {\em Journal of Artificial Societies and Social Simulation}, 7(3),
  2004.

\bibitem{suzuki}
J~Suzuki.
\newblock A {M}arkov chain analysis on simple genetic algorithms.
\newblock {\em IEEE Transactions on Systems, Man, and Cybernetics},
  25:655--659, 1995.

\bibitem{mspaper5ThomasSycara1998}
James Thomas and Katia Sycara.
\newblock Heterogeneity, stability, and efficiency in distributed systems.
\newblock In Yves Demazeau, editor, {\em International Conference on Multi
  Agent Systems}, pages 293 -- 300. IEEE Press, 1998.

\bibitem{vallee1995cognition}
R.~Vallee.
\newblock Cognition et systeme (cognition and systems).
\newblock {\em Paris: l'Interdisciplinaire Systeme (s)}, 1995.

\bibitem{weiss1999msm}
G.~Weiss.
\newblock {\em Multiagent Systems: A Modern Approach to Distributed Artificial
  Intelligence}.
\newblock MIT Press, 1999.

\bibitem{weisstein2003cce}
E~Weisstein.
\newblock {\em {CRC} Concise Encyclopedia of Mathematics}.
\newblock CRC Press, 2003.

\bibitem{wiener1948cybernetics}
N.~Wiener.
\newblock {\em Cybernetics or Control and Communication in the Animal and the
  Machine}.
\newblock MIT Press, 1948.

\bibitem{wooldridge}
M~Wooldridge.
\newblock {\em Introduction to MultiAgent Systems}.
\newblock Wiley, 2002.

\bibitem{wright1932}
S~Wright.
\newblock The roles of mutation, inbreeding, crossbreeding and selection in
  evolution.
\newblock In D.~Jones, editor, {\em International Congress on Genetics}, pages
  356--366. Brooklyn botanic garden, 1932.

\end{thebibliography}

\vfill

\section*{Biographies}

\vspace{-10mm}

\begin{biography}[{\includegraphics[width=1in]{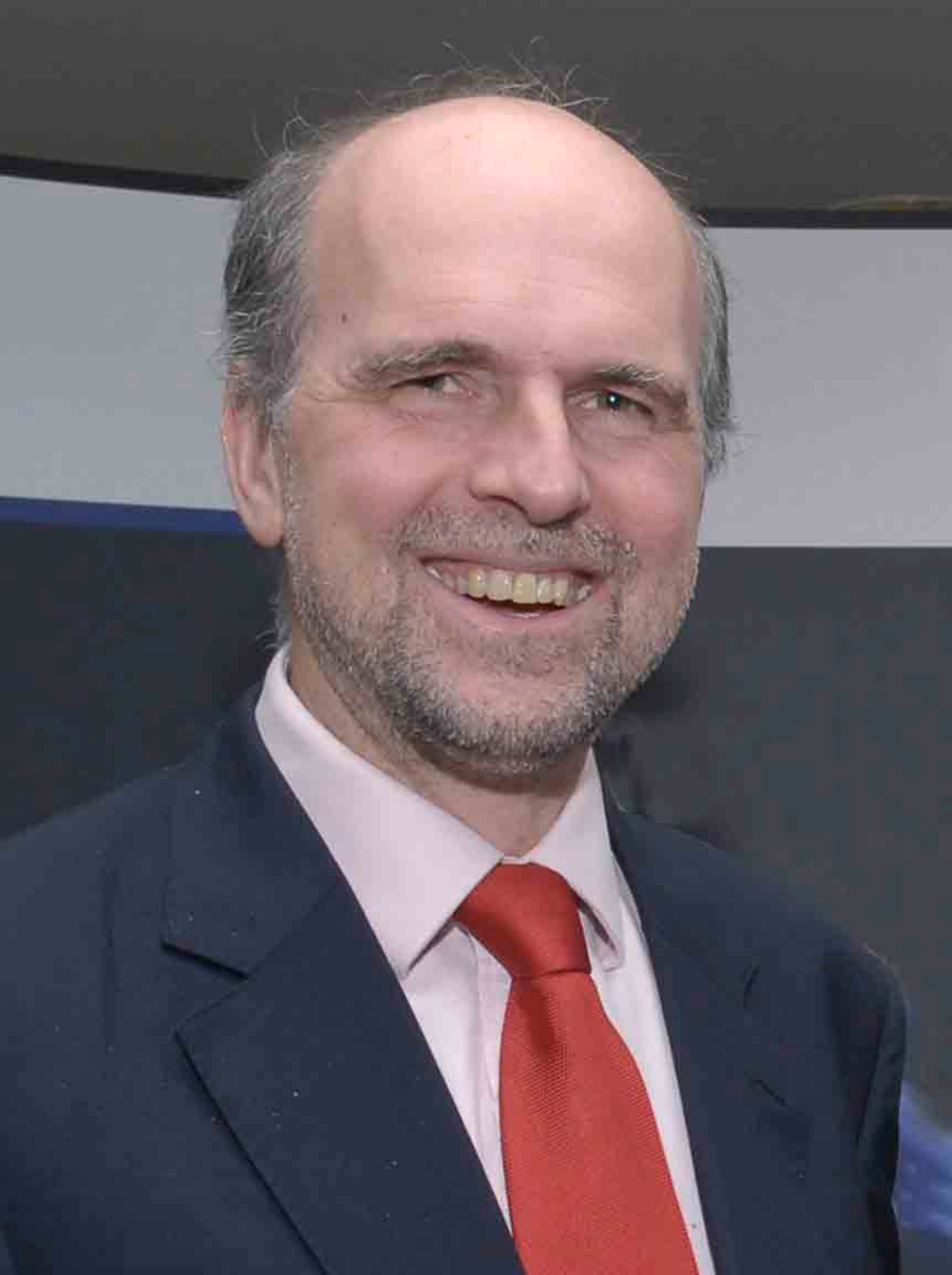}}]{Philippe De Wilde} is a Professor at the Intelligent Systems Lab, Department of Computer Science, and Head of the School of Mathematical and Computer Sciences, Heriot-Watt University, Edinburgh, United Kingdom. Research interests: stability, scalability and evolution of multi-agent systems; networked populations; coordination mechanisms for populations; group decision making under uncertainty; neural networks, neuro-economics. He tries to discover biological and sociological principles that can improve the design of decision making and of networks. Research Fellow, British Telecom, 1994. Laureate, Royal Academy of Sciences, Letters and Fine Arts of Belgium, 1988. Senior Member of IEEE, Member of IEEE Computational Intelligence Society and Systems, Man and Cybernetics Society, ACM, and British Computer Society. Associate Editor, IEEE Transactions on Systems, Man, and Cybernetics, Part B, Cybernetics.
\end{biography}

\vspace{-10mm}

\begin{biography}[{\includegraphics[width=1in]{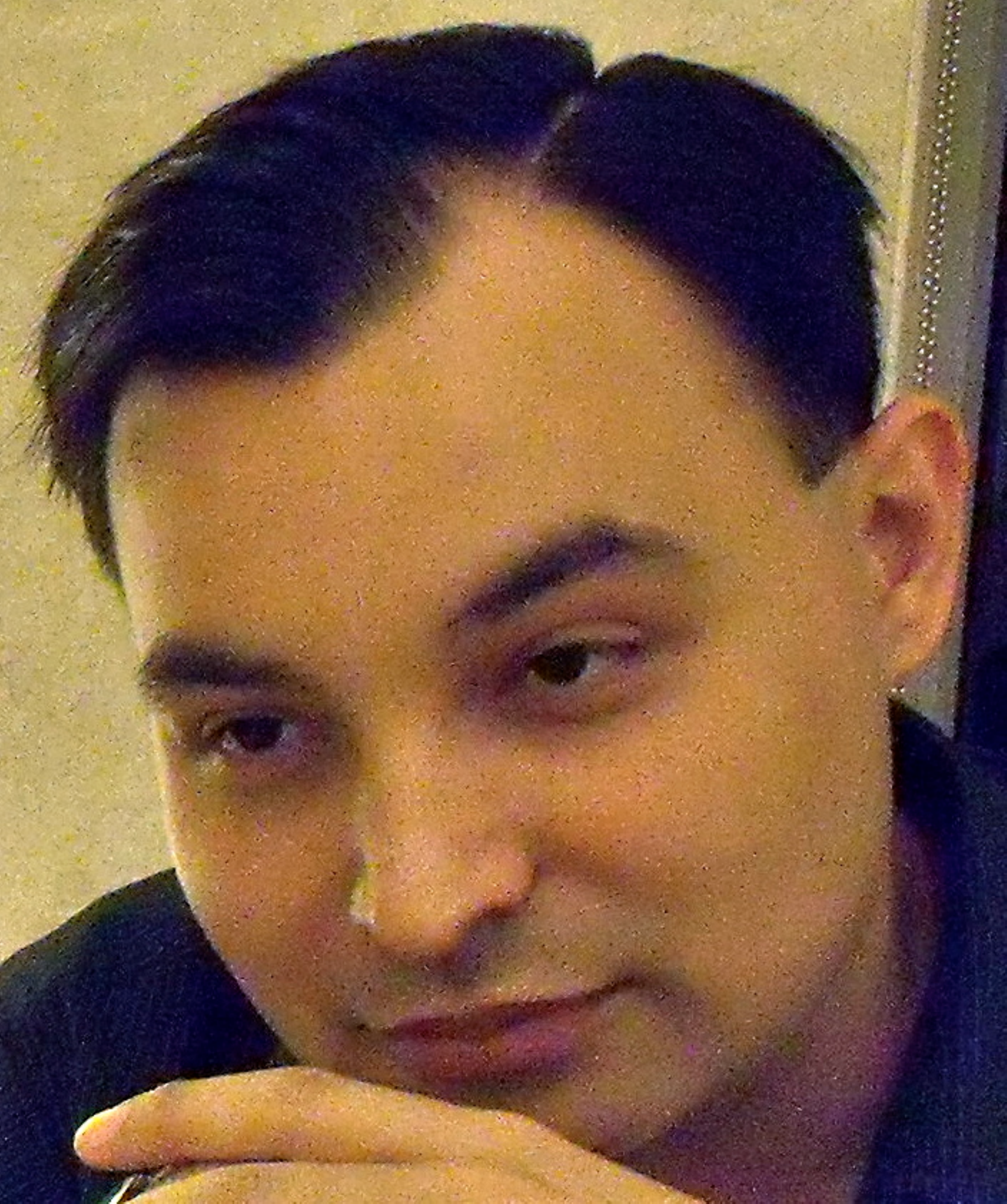}}]{Gerard Briscoe} is a Research Associate at the Systems Research Group of the Computer Laboratory, University of Cambridge, UK, and a Visiting Scholar at Intelligent Systems Lab of the School of Mathematical and Computer Sciences, Heriot-Watt University, UK. Before this he was a postdoctoral researcher at the Department of Media and Communications of the London School of Economics and Political Science, UK. He received his PhD in Electrical and Electronic Engineering from Imperial College London, UK. Before which he worked as a Research Fellow at the MIT Media Lab Europe, after completing his B/MEng in Computing also from Imperial College London. His research interests include Sustainable Computing, Cloud Computing, Social Media and Natural Computing.
\end{biography}

\vfill

\end{document}